\documentclass[aps,showpacs,twocolumn,floatfix,pra,superscriptaddress]{revtex4}
\usepackage{graphicx}
\usepackage{amsfonts}
\usepackage{amsmath}
\usepackage{amssymb}
\usepackage[usenames,dvipsnames]{color}

\begin{document}
\title{Perturbation spreading in many-particle systems: \\ a random walk approach}
\author{V. Zaburdaev}
\affiliation{School of Engineering and Applied Science, Harvard University,
29 Oxford Street, Cambridge, Massachusetts 02138, USA}
\author{S. Denisov}
\author{P. H\"{a}nggi}
\affiliation{ Institute of Physics, University of
Augsburg, Universit\"{a}tsstrasse~1, D-86159 Augsburg, Germany}

\begin{abstract}

The propagation of an initially localized perturbation via an interacting many-particle Hamiltonian dynamics is investigated. We argue that the propagation of the perturbation can be captured by the use of a continuous-time random walk where a single particle is traveling through an active, fluctuating medium. Employing two archetype ergodic many-particle systems, namely (i) a hard-point gas composed of two unequal masses and (ii) a Fermi-Pasta-Ulam chain we demonstrate that the corresponding perturbation profiles coincide with the diffusion profiles of the single-particle L\'{e}vy walk approach. The parameters of the random walk can be  related through elementary algebraic expressions to the physical parameters of the corresponding test many-body systems.
\end{abstract}
\maketitle

The transport properties of many-particle systems are of salient interest in diverse contexts, ranging from foundations of thermodynamics to the transduction of information on the nanoscale. The collective evolution of   $N \gg 1$ interacting particles creates a dynamical ``tissue,'' whose properties depend not only on the Hamiltonian of the system,  but also on the state of the system itself. An objective of primary interest is how the system responds to the perturbation that locally affects its dynamics.
The answer then provides  direct insight into collective energy, correlation, and information transport in extended nondissipative media  \cite{old, torc, primo, politi, cipr, wang, flach}.

Consider the situation of a many-particle system at microcanonical equilibrium, when at the initial time $t=0$ one of the particles receives some external local perturbation. The system gains a small amount of \textit{perturbation energy}, which is conserved due to Hamiltonian evolution of the system. However, the perturbation does spread as the perturbation energy is shared by a constantly growing number of particles.
\begin{figure}[t]
\center
\includegraphics[width=8.4cm]{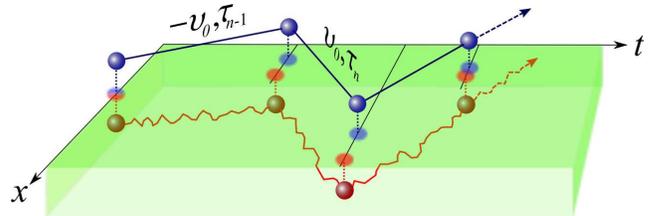}
\caption[Models]
{(color online) The standard continuous-time random walk (above the slab) and random walk through an active medium (in the slab). The first walker performs flights with constant velocity $\pm v_0$, while the second one is subjected to constant scattering, so that its velocity fluctuates. The duration of single flight, $\tau_n$, is a random variable drawn from the probability density function $\psi(\tau)$.}
\label{fig1}
\end{figure}
One of the main features of the spreading is the {\it finite} velocity of perturbation propagation, $v_{0} < \infty$ \cite{politi, wang, flach}.  Therefore, similar to relativistic diffusion theories \cite{Dunkel}, an effective ``light cone'' occurs \cite{cone}, such that  at the given time $t$ the  perturbation is almost completely confined to the interval $[ -v_{0} t, v_{0} t]$ \cite{almost}. The fundamental fact of the cone's existence assumes the characteristics of a mathematical existence theorem \cite{cone, lrb}; its strength therefore is the generality when dealing with  many-particle systems in rather arbitrary situations. A pronounced weakness of this mathematical approach, however,  becomes evident whenever one attempts to implement the theory for a particular system to obtain  qualitative results on an analytical level of description. Therefore, more applied approaches, like the diffusion formalism, might become of beneficial use. Indeed it is appealing to consider the said perturbation spreading  as a certain (not yet known) one-dimensional diffusion process, and to quantify it with the mean square displacement and other related attributes \cite{old, cipr, wang, flach}. As it is known, however, it is impossible to describe this process by any known macroscopic, norm-preserving normal diffusion equation. Conventional diffusion equations knowingly lead to infinite propagation speeds \cite{Dunkel} and therefore are incompatible with the existence of a causal cone. Thus the perturbation kinetics should be ultimately considered on the microscopic level corresponding to the random walk approach.

With this Letter we employ the microscopic single-particle random walk process in order to evaluate the evolution of perturbations in one-dimensional, ergodic  many-particle systems. In doing so we  address the two challenges:  (i) How is the perturbation distributed within the cone, and (ii) what are  the shapes of the cone fronts? %First, we introduce a random walk model for a walker submerged into the fluctuating medium, cf. Fig. 1. We extend the continuous-time random walk (CTRW) formalism \cite{ctrw} to include the ``walker-media'' interactions.
By using two renowned many-particle chains we demonstrate that the random walk model accurately describes the perturbation spreading in these {\it interacting} many-body systems. We  show that the walker-media interactions are responsible for the observed shape of the causal cone and predict the universal scalings for the perturbation profile and its corresponding fronts.

\begin{figure}[t]
\center
\includegraphics[width=0.34\textwidth]{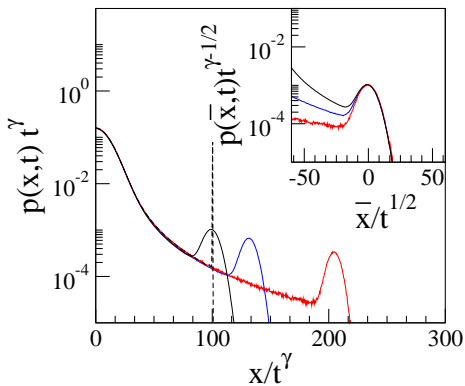}
\caption[]
{(color online) Rescaled propagators of  L\'{e}vy walk process with exponent $\gamma = 5/3$ for different times, $t = 100, 300, 600$. The dashed line depicts the propagator for the standard L\'{e}vy walk  with the constant flight velocity $v_0 = 1$. For the simulations of the random walk in the active medium we employed the map (\ref{eq:fluct}) with  $D_{v} = 0.03$. The inset shows the ballistic front regions after the scaling transformation (\ref{eq:scaling2}), where $\bar{x} = x - v_0 t$.}
\label{fig2}
\end{figure}

\textit{Model setup.}
The continuous-time random walk (CTRW) formalism \cite{ctrw} has found applications to a wide range of phenomena, ranging from financial markets dynamics \cite{market} to  single molecule spectroscopy \cite{molecule}. Here, we use one specific CTRW model \cite{sher, klafter1}, where a walker moves ballistically in between successive ``turning points''. During a single flight event, the walker travels at constant speed $v_{0}$,  and at the turning points it randomly changes the direction of motion (see the particle flying above the slab on Fig. 1). Hence,  the velocity probability density function (PDF) reads $h(v)=\left[\delta(v-v_{0})+\delta(v+v_{0})\right]/2$. The flight times, $\tau_n$, are independent and identically distributed random variables drawn from a PDF $\psi(\tau)$ that is  described by a power law \cite{cipr, klafter1, levy, klafter2, klafter3},
\begin{equation}
\psi(\tau) \propto (\tau/\tau_{0})^{-\gamma-1},
\label{eq:pdf}
\end{equation}
where $\tau_{0}$ is a characteristic time scale, and $1 < \gamma \leq 2$. This choice  guarantees a finite average flight time, $\left<\tau\right>=\int_{0}^{\infty} \tau \psi(\tau) d\tau $, and  provides access to different diffusion regimes with the scaling of the mean squared displacement  $\sigma^{2}(t) \propto t^{3 - \gamma}$. The corresponding L\'{e}vy walk (LW) approach \cite{klafter1} has been successfully used for the description of  diffusion of particles in chaotic systems \cite{levy, klafter2, klafter3}, tracers in turbulent flows \cite{solomon}, or ultracold atoms in optical potentials \cite{cold}.  For the standard LW process,  the PDF of finding a particle in $x$ at the time $t$, $P(x,t)$,  provided it was initially localized at $x=0$,  exhibits a sharp cutoff marked by the ballistic peaks at $|x|=v_0 t$ ~\cite{klafter3}.

A first step towards the implementation of the LW formalism for the description of the perturbation dynamics in many-particle systems has been attempted in \cite{cipr}. Although this approach  provided an adequate description for the perturbation spreading process, it failed to capture the dynamics of the cone fronts. The observed ballistic humps displayed  smooth, Gaussian-like profiles, with a scaling that was incompatible with the scaling behavior of deltalike peaks of the  LW propagator \cite{klafter2}.

In order to resolve this issue, we here  extend the conventional CTRW setup by assuming that the walker performs a random walk through an {\it active} medium. Conceptually, it means that while moving, the walker interacts with the surrounding medium (the slab in Fig. 1). This interaction causes fluctuations of the walker's velocity (note the noisy trajectory of the particle moving inside the slab in Fig. 1). The term ``active'' means that the medium is not solely dissipative and the particle not only continuously loses but also gains energy from its environment. Both processes are in balance, yielding  unbiased fluctuations of the walker's velocity around $v_0$.

\textit{Model dynamics.} We start out by considering a single flight event. The position of the walker is defined by a Langevin equation
%\begin{equation}
$\dot{x}=v_0+\xi(t)$,
%\end{equation}
where $\xi(t)$ is a delta-correlated Gaussian process of vanishing mean and  finite intensity $D_v$, i.e.,  $\left<\xi(t) \xi(s)\right>=D_v\delta(t-s)$. This constitutes a well-known biased Wiener process with drift $v_0$ \cite{kartas}. After an integration over a time interval $\tau$ we obtain
\begin{equation}
x(t+\tau)=x(t)+v_0\tau+w(\tau),
\label{eq:fluct}
\end{equation}
where  the new stochastic variable $w(\tau)=\int_{t}^{t+\tau}\xi(s)ds$ is characterized by the Gaussian PDF $p(w,\tau)$ with the dispersion $\sigma_{\tau}^2=\langle [x(\tau) - v_0\tau]^2\rangle = D_v\tau$.
%Therefore if the walker starts its flight of duration $\tau$ at $x$ with velocity $\upsilon$, it will end up at %$x+v\tau+w(\tau)$, where $w$ is a Gaussian random variable with the pdf $p(w,\tau)$.
The propagator can be calculated numerically by sampling  long enough a single-particle trajectory \cite{initial}, (see Fig. 2).

To gain analytical insight into the generalized LW dynamics, we follow a standard reasoning \cite{klafter2,zaburdaev} and derive the transport equations governing the evolution of the particle density, $P(x,t)$ (see supplemental material). Below we present the major results for the scaling properties of the central part of the density profile and provide the explicit expression describing the ballistic humps.  The asymptotic analysis of the central part of the density profile reveals the scaling of the standard LW propagator \cite{klafter2}, namely,
%
% Here I changed "s" into "u" in order to be consistent with notation in supplementary material
%
\begin{equation}
P(x,t')\simeq \frac{1}{K u^{1/\gamma}} P \left(\frac{x}{K u^{1/\gamma}},t\right),~~~~ |x| \ll v t
\label{eq:scaling1}
\end{equation}
where $K \propto \tau_{0}^{1-1/\gamma}v_0$ and $u = t'/t$, see Fig. 2.

The salient difference between the dynamics of our model and the standard L\'{e}vy walk becomes apparent in the regions of cone fronts. The ballistic humps are formed by the particles which were flying from $t=0$ to the observation time $t$ (see supplemental material):
\begin{equation}
P_{\text{hump}}(x,t)=\Phi(t)\left[p(x+v_0t,t)+p(x-v_0t,t)\right]/2
\label{n_hump}
\end{equation}
Here, $\Phi(t)$ denotes the probability of not changing the direction of flight during the time $t$ and has a power-law behavior $\Phi(t)\propto(t/\tau_{0})^{1-\gamma}$ \cite{klafter2}. Consequently, the area under the ballistic humps (\ref{n_hump}) also scales as $t^{1-\gamma}$. During flights, the particles undergo random fluctuations caused by the flight's velocity variations. The flight length is proportional to $t$; thus the dispersion of
the Gaussian-like humps grows as $\sqrt{t}$, and we arrive at the following scaling for the particles' density in the
hump regions:
\begin{equation}
P_{\text{hump}}(\bar{x},t') \simeq u^{-1/2}P_{\text{hump}}(\bar{x}/u^{\gamma-1/2},t),
\label{eq:scaling2}
\end{equation}
where $u = t'/t$ and $\bar{x} = x - v_0 t$ (see inset in Fig. 2). Note that this scaling  distinctly differs from the scaling  in  Eq. (\ref{eq:scaling1}).

\textit{Ergodic  many-body systems: validation of the approach.} Consider a many-particle system, with a Hamiltonian
\begin{equation}
H_{\text{total}}(\{x_i,p_i\} ) = \sum^{N}_{i=1} H_i,
\label{eq:total}
\end{equation}
where $H_i = H(x_i, x_{i-1}, x_{i+1}, p_i)$ is the energy attributed to the $i$th particle. At
the time $t=0$ the system is locally affected by the perturbation.
%, which changes the dynamics of a single particle, or few neighbor particles,  only.
%Denote the perturbation energy by $E_p$.
% At the time $t=0$ it is locally affected by a perturbation, which changes the dynamics of a single particle, or few neighbor particles,  only.
%Denote the perturbation energy by $E_p$.
The initially localized perturbation energy, $E_p$, starts to spread, such that the distribution of the local excess energy $\triangle E(i,t)$ of the $i$ particle evolves in time \cite{ex}, while keeping the perturbation  energy constant; i.e., $\Sigma_{i=1}^{N} \triangle E(i,t) =  E_p$. The spreading can be quantified with a normalized probability distribution function $\varrho (i,t)= \overline{\triangle E(i,t)}/ E_p$, where $\overline{{\cdot \cdot \cdot}}$ denotes a microcanonical average. %This MSD of the spreading process is defined as $\sigma^{2}(t)= \Sigma_{i=1}^{N} i^{2} \varrho (i,t)$ \cite{old, cipr, wang, flach}.

The main finding of this study is that the profiles of the spreading perturbation in many-particle Hamiltonian systems, $\varrho (i,t)$, in the corresponding asymptotic regime \cite{time}, are determined by the propagator $P(x,t)$ of the generalized L\'{e}vy walk model, Eqs. (\ref{eq:scaling1}, \ref{eq:scaling2}). In order to validate this claim we use two archetype systems, namely (i) a one-dimensional scattering dynamics of a hard-point gas  composed of two unequal masses \cite{hpg} and (ii) a Fermi-Pasta-Ulam (FPU) $\beta$-lattice dynamics \cite{fpu}.

\begin{figure}[t]
\center
\includegraphics[width=0.48\textwidth]{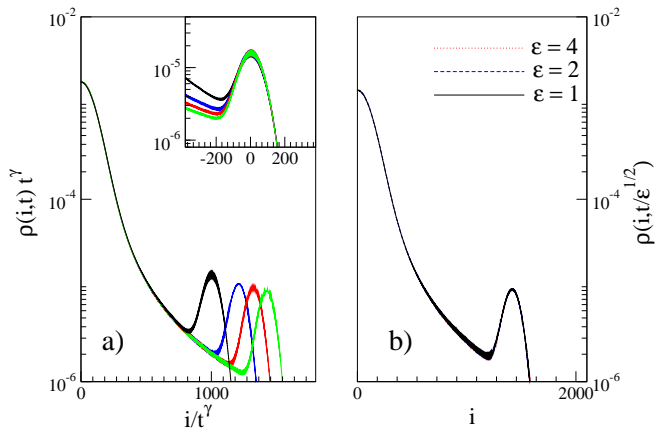}
\caption[]
{(color online) (a) Rescaled perturbation profiles at times $t=1000, 2000, 4000$, and $6000$ (the width increases with time), for the hard-point gas dynamics with the energy per  particle $\varepsilon = 1$. The scaling exponent is $\gamma = 5/3$. The inset, similar to the inset in Fig. 2, depicts the ballistic humps after the scaling transformation (\ref{eq:scaling2}). (b) Profiles after the scaling transformation (\ref{eq:scaling_hpg}) for different energy per particle, $\varepsilon$, at time $t=1500/\sqrt{\varepsilon/\varepsilon_0}$, with $\varepsilon_0 = 1$. Each profile is obtained by averaging over $10^{6}$ realizations.}
\label{fig3}
\end{figure}

We start with a hard-point gas, a many-body Hamiltonian system with an ergodic dynamics governed by the conservation of kinetic energy and momentum \cite{hpg}. We use a chain of $N = 1.6 \cdot 10^{4}$ pointlike particles with alternating masses, $...m M m M...$, of the length $L=1.6 \cdot  10^{4}$, and periodic boundary conditions. Without loss of generality we set the mass ratio $M/m=2$. The energy per particle, $\varepsilon = \langle m_{i} v^{2}_i\rangle/2$, $m_i = m$ or $M$, serves as a tunable parameter. Figure 3 depicts the evolution of the infinitesimal perturbation $\varrho (i,t)$  \cite{cipr}. The scaling ansatz (\ref{eq:scaling2}) with the exponent $\gamma = 5/3$ is beautifully validated, see the inset in  Fig. 3a. We also found that the perturbation profiles for different values of energy per particle parameter $\epsilon$ are matched by assuming that the perturbation velocity and the fluctuation variance  both scale as
\begin{equation}
v_0, D_\upsilon \propto \sqrt{\varepsilon}.
\end{equation}
Consequently,
%\begin{equation}
%\upsilon_{m} \propto \sqrt{\epsilon},
%\end{equation}
the profile scales as
\begin{equation}
\varrho_{\varepsilon}(x,t) = \varrho_{\varepsilon'}(x,t/ s'),
\label{eq:scaling_hpg}
\end{equation}
where $s' = \sqrt{\varepsilon'/\varepsilon}$, see Fig. 3b.

As our second test bed we use a FPU $\beta$ chain dynamics \cite{fpu}, defined by the Hamiltonian (\ref{eq:total}) with
$H_i = \frac{1}{2}p_{i}^{2} +\frac{1}{2} (x_{i+1} - x_{i})^{2}$ $+  \frac{\beta}{4} (x_{i+1} - x_{i})^{4}$,
with $N$ particles of unit mass and periodic boundary conditions. The energy per particle is $\varepsilon = H_{\text{total}}/N$. It is  not feasible to explore the evolution of  finite perturbations of the FPU system at microcanonical equilibrium, due to emerging  huge statistical fluctuations.
%because of huge statistical fluctuations which prevent from obtaining more or less sufficient statistics.
Instead we employed the energy correlation function, $e(i,t)$ \cite{wang, flach}, which  bears the same information as the infinitesimal perturbation in the case of hard-point gas \cite{detail}. We performed a massive numerical experiment \cite{supp1},
%which allowed a remarkable speedup due to high parallelization of computations.
yet even these efforts were not sufficient to cope with the statistical fluctuations [note the thin green lines in Figs. 4a,b]. Nevertheless,  a relatively smooth shape of the ballistic hump allows for a convincing validation of the scaling (\ref{eq:scaling2}).
It is interesting to note that the velocity of the ballistic peaks is determined by the group velocity of effective thermal phonons \cite{flach}, which, therefore can be associated with ''walkers'' of the CTRW approach.
\begin{figure}
\center
\includegraphics[width=0.46\textwidth]{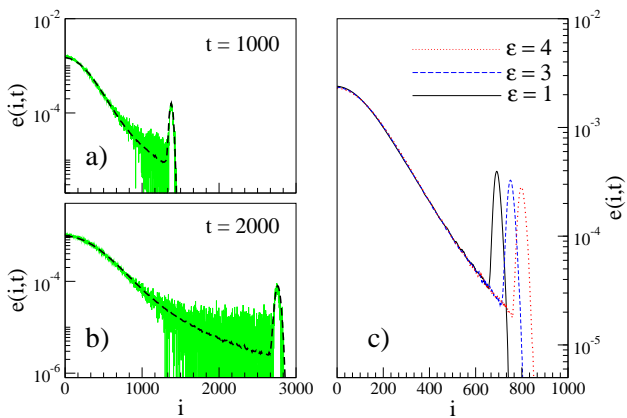}
\caption[]
{(color online) (a, b) Energy correlation functions $e(i,t)$ for the FPU dynamics  with $N = 1.6 \cdot 10^{4}$, $\beta = 1$ and $\varepsilon = 1$ (thin solid lines), compared to the propagators of the generalized LW model with $v_0 = 1.384$ and $D_{v} = 0.49$  (thick dashed lines). The scaling exponent  is $\gamma = 5/3$.
The FPU profiles have been obtained by averaging over  $5 \cdot  10^{5}$  realizations. (c)   Energy  correlation functions for the FPU system  with $N = 2 \cdot 10^{3}$ for the time $t=500$ and different values of the energy per particle $\varepsilon$.}
\label{fig4}
\end{figure}

The perturbation profiles for different values of  the energy per particle $\varepsilon$ reveal another remarkable feature: the central part of profiles is independent of $\varepsilon$, while the ballistic humps move   as $\varepsilon$ increases, see Fig. 4c. Such behavior can be derived from the scaling invariance (\ref{eq:scaling1}). Taking into account that  $K \propto \tau_{0}^{1-1/\gamma}v_0$,  one can infer that the central part of the LW propagator is invariant under variation of $v_0$ when $\tau_0 \propto v_0^{\gamma/(1-\gamma)}$.
% yet the refinement of this connection still remains an challenging problem.

%By knowing the dependence $v_{0}(\epsilon)$, one could derive then the relation between the mean free propagation time %of thermal phonons, $\tau_{0}$,  and the microcanonical temperature,  $\epsilon$.
In conclusion, we demonstrated that the collective process of perturbation spreading across two celebrated \textit{many-particle} systems, a hard-point gas and a Fermi-Pasta-Ulam lattice, is reproduced by a \textit{single-particle} stochastic process. It is intuitive that an
ergodic dynamics is a prerequisite for the diffusionlike perturbation evolution. In integrable many-body systems, such as a harmonic chain or the Toda lattice \cite{toda}, the perturbation spreading is a deterministic process, evolving in terms of  noninteracting phonons. The task of exploring the sufficient conditions for the CTRW kinetics to occur presents a promising challenge, thereby underpinning the universality of our findings. Thus our results   disclose a pathway to explore  propagation of information in realistic dissipation-free systems: it allows to calibrate the transport characteristics of many-body systems (which are  beyond  the region of validity of the classical Lieb-Robinson theory \cite{lrb})  by using  the parameters of the corresponding random walk process.

Apart from those theoretical challenges, there is room for  possible applications. A realization that comes to mind is an array of coupled nanoresonators \cite{nano},  where a single unit acts as both, the receiver and transducer of excitations, which  transforms this array into an extended sensor via utilizing the principle of time-of-arrival localization \cite{acu}.

This work has been supported by the DFG Grants No. HA1517/31-2 (S. D. and P. H.) and No. ZA593/2-1 (V.Z.).

\newpage

\title{Supplemental Material for \\`Perturbation spreading in many-particle systems: a random walk approach}
\maketitle

\section{Transport equations for the model propagator}

In order to gain analytical insight into the generalized LW dynamics, we proceed along known reasoning, see in   \cite{klafter2,zaburdaev}, and derive the transport equation for the probability distribution function (PDF) $P(x,t)$.  We first introduce the probability distribution of the end points of flights, or the turning points, $\nu(x,t)$, where a particle chooses its new velocity. Its evolution is governed by the following balance equation: \begin{widetext}

\begin{equation}
\nu(x,t)=\int\limits_{-\infty}^{\infty}\int\limits_{-\infty}^{\infty}dvdw\int\limits_{0}^{t}\nu(x-v\tau-w,t-\tau)\psi(\tau)h(v)p(w,\tau)d\tau
+\varphi(t)\int\limits_{-\infty}^{\infty}h(v)p(x-vt,t)dv\label{eq:nu}\;.
\end{equation}
\end{widetext}

Equation (\ref{eq:nu}) shows that a particle changes its velocity at the point $(x,t)$ if it was the end point of the preceding step. That previous step had some flight time $\tau$ and occurred with some velocity $v$. Therefore the step originated in the point  $x-v\tau-w$, where $w$ takes into account the accumulation of velocity fluctuations during a single flight.  Next we have to integrate over all possible flight times, velocities and fluctuations with the corresponding probability densities. This is how the first term on the right side of Eq. (\ref{eq:nu}) is obtained. The second term reflects the influence of initial conditions. In this work we use so-called equilibrated initial conditions \cite{tunaley} which assumes that walkers evolved for an infinitely long time when the observation started. Different initial setups affect only the probability of when a flying particle experiences the first turn after the start of observation. For a system evolving for an infinite time the PDF of the first turn after the observation has been started is given by  $\varphi(t)=\left<\tau\right>^{-1}\int_{0}^{\infty}\psi(t+\tau)d\tau$ \cite{tunaley}. If a particle starts at $x=0$, the spatial position of where the first turn occurs is influenced by the velocity fluctuations and is given by: $\int_{\infty}^{\infty}\delta(x-vt-w)p(w,t)dw=p(x-vt,t)$.

We next   evaluate the actual density of particles $P(x,t)$, to obtain:
\begin{widetext}
\begin{equation}
P(x,t)=\int\limits_{-\infty}^{\infty}\int\limits_{-\infty}^{\infty}dvdw\int\limits_{0}^{t}\nu(x-v\tau-w,t-\tau)\Psi(\tau)h(v)p(w,\tau)d\tau
+\Phi(t)\int\limits_{-\infty}^{\infty}h(v)p(x-vt,t)dv \;.
\label{eq:n}
\end{equation}
\end{widetext}
The particle currently finds itself in the point $(x,t)$, if it previously changed the direction of flight at the point $x-v\tau-w$, and keeps flying afterwards for the time $\tau$ with the probability $\Psi(\tau)=1-\int_{0}^{\tau}\psi(t)dt$.  Similarly,  the probability to continue the first flight reads $\Phi(t)=\left<\tau\right>^{-1}\int_{0}^{\infty}\Psi(t+\tau)d\tau$ \cite{klafter2}.

Formally, the above two equations can be solved with a help of combined Fourier- and Laplace-transform, thereby turning all convolution integrals into  products and thus rendering the integral equations algebraic. The answer for the density of particles in Fourier/Laplace space, $\widetilde{P}_{k,s}$ is given by (we use tilde-notation to denote the Fourier/Laplace transform):
\begin{equation}
\widetilde{P}_{k,s}=
\frac{\left[\Psi(\tau)\widetilde{h}_{k\tau}\widetilde{p}_{k}(\tau)\right]_s\left[\varphi(\tau)
\widetilde{h}_{k\tau}\widetilde{p}_{k}(\tau)\right]_s}{1-[\widetilde{h}_{k\tau}\widetilde{p}_{k}
(\tau)\psi(\tau)]_{s}}+\left[\widetilde{h}_{kt}\widetilde{p}_{k}(\tau)\Phi(t)\right]_s \;.\label{density_answ}
\end{equation}
This exact analytical expression  serves as the starting point for the asymptotic analysis for large spatial and temporal scales, that corresponds to small $k, s$ coordinates in Fourier/Laplace-space. It is possible to show that for the case of Gaussian fluctuations the central part of the profile can be described by the L\'{e}vy distribution, which is  the solution of the fractional diffusion equation \cite{klafter1,klafter2}:
$$\widetilde{P}_{k,s}\simeq\frac{1}{s+\tau_{0}^{\gamma-1}v_0^{\gamma}(\gamma-1)\Gamma[1-\gamma]k^{\gamma}\sin(\pi\gamma/2)}\;\;\;. $$
In original coordinate space and original time this expression delivers the following scaling relation for the central part of the density profile:

\begin{equation}
P(x,t')\simeq \frac{1}{K u^{1/\gamma}} P \left(\frac{x}{K u^{1/\gamma}},t\right),~~~~ |x| \ll v t \;,
\label{eq:scaling1}
\end{equation}
where $K \propto \tau_{0}^{1-1/\gamma}v_0$ and $u = t'/t$, see Fig. 2.

To describe the density of particles in the ballistic humps of the profile we use the second term on the right hand side of Eq. (\ref{eq:n}):

\begin{equation}
P_{\text{hump}}(x,t)=\Phi(t)\int\limits_{-\infty}^{\infty}h(v)p(x-vt,t)dv \;,
\label{hump}
\end{equation}
where  $\Phi(t)=\left<\tau\right>^{-1}\int_{0}^{\infty}\Psi(t+\tau)d\tau$. For the chosen flight time distribution it scales as $\Phi(t) \propto (t/\tau_0)^{1-\gamma}$.

\section{Numerics}

For the integration of the FPU $\beta$ chain's equations  of motion we used the symplectic $SABA_{2}C$ scheme \cite{sabac1, sabac2}, with the integration time step $dt = 0.01 \div 0.02$.  Calculations have been performed on a Tesla S1070 supercomputer, with 960 CPU's on board.


\begin{thebibliography}{99}
\bibitem{almost} The term ``almost''  underpins the fact that outside the cone the perturbation is strongly diminished.
\bibitem{initial} This setup amounts to equilibrated initial conditions, when at time $t=0$ the particle is in the flight with probability one. It should be distinguished from nonequilibrated conditions when at the time $t=0$ the particle is set at rest [15].
\bibitem{ex}  This quantity can be computed as follows: at time $t=0$ the system is locally perturbed, $\{x^{\text{p}}_{0}(0)=x_{0}(0)+ \epsilon_x, p^{\text{p}}_0(0)=p_0(0)+\epsilon_p\}$. Then the two systems, original and perturbed, evolve in parallel. The local excess  of energy is given by $\triangle E(i,t)= H^{\text{p}}_i(t) - H_i(t)$.
\bibitem{time}  Practically, this implies the existence of a time scale, $t_{LW}$, beyond which all relevant scattering mechanisms  unfold themselves. This time $t_{LW}$ is system specific.
\end{thebibliography}
\end{document}